\documentclass[twocolumn,prb,preprintnumbers,amsmath,amssymb]{revtex4-1}

\usepackage{graphicx}
\usepackage{dcolumn}
\usepackage{bm}

\begin{document}


\title{Effect of potential fluctuations on shot noise suppression in mesoscopic
cavities}

\author{P. Marconcini}
\author{M. Totaro}
\altaffiliation[Current Affiliation: ]
{Center for Microbiorobotics @SSSA, Istituto Italiano di Tecnologia,
Viale Rinaldo Piaggio 34, 56025 Pontedera (PI), Italy.}
\author{G. Basso}
\author{M. Macucci}
\affiliation{Dipartimento di Ingegneria dell'Informazione,\\
Universit\`a di Pisa, Via Girolamo Caruso 16,
56122 Pisa, Italy.}

\begin{abstract}
We perform a numerical investigation of the effect of the disorder associated
with randomly located impurities on shot noise in mesoscopic
cavities.
We show that such a disorder 
becomes dominant in determining the noise behavior when the amplitude
of the potential fluctuations is comparable to the value of
the Fermi energy and for a large enough density of impurities. In contrast to 
existing conjectures, random potential fluctuations are shown not to 
contribute to achieving the chaotic regime whose signature is a Fano factor of 
1/4, but, rather, to the diffusive behavior typical of disordered conductors. 
In particular, the 1/4 suppression factor expected for a symmetric cavity 
can be achieved only in high-quality material, with a very
low density of impurities. As the disorder strength is increased, a 
relatively rapid transition of the suppression factor from 1/4 to values 
typical of diffusive or quasi-diffusive transport is observed. Finally, on 
the basis of a comparison between
a hard-wall and a realistic model of the cavity, we conclude that the 
specific details
of the confinement potential have a minor influence on noise.
\end{abstract}

\pacs{73.50.Td, 85.35.Be, 73.50.-h}
\keywords{chaotic cavity, disorder, shot noise, diffusive transport}
\maketitle

\section{Introduction}
In the last decades, the rapid development of nanofabrication techniques for
sub-micron devices has led to impressive advancements in the field of
mesoscopic physics, with a much better understanding of the properties of 
low-dimensional conductors. Molecular beam epitaxy (MBE) has been perfected
to the point of growing extreme high-quality 
heterostructures~\cite{radu,umansky}, exhibiting low-temperature mobilities, 
for the case of a 2DEG in the AlGaAs/GaAs material system, up to $3.5 \times 
10^7 $~cm$^2$V$^{-1}$s$^{-1}$ (see Ref.~\cite{umansky}). These 
heterostructures represent ideal testbeds for the study of the quantum 
properties of confined electrons.

A field of research that has become very active in recent years is the 
investigation of the noise properties of mesoscopic devices. In particular, 
shot noise has been the subject of significant attention~\cite{blantbuett,
heiblum,revdejong,been,henny,jala,ober1,natpap,sercav,jcirc,dejong2,serbar}, 
because it can 
provide useful information about material and device properties. 

The fundamental nature of shot noise, resulting from the discreteness of 
charge, was recognized already at the beginning of last century, when 
Schottky formulated his well-known theorem~\cite{Schottky}, according 
to which, in the absence of correlation between carriers, the shot noise 
power spectral density $S_I$ is given by the expression 
$S_I=2e\langle
I \rangle$, where $e$ is the electron charge and $\langle I \rangle$
the average current in the conductor. This is a result of the Poissonian 
statistics governing device traversal events. In the presence of correlations 
among carriers the shot noise power spectral density does differ from
Schottky's result. Especially in mesoscopic
devices, strong correlations, mainly due to Pauli exclusion
and to Coulomb interaction, lead to considerable deviations from the Poissonian
limit.  


The so called Fano factor
\begin{equation}
F = \frac{S_I}{S_{full}}=\frac{S_I}{2e\langle I \rangle},
\end{equation}
is defined as the ratio of the actual shot noise power spectral density 
$S_I$ to that which would be expected from Schottky's formula. 

In the Landauer-B\"uttiker formalism the Fano factor can be
expressed as~\cite{buettprl}: 
\begin{equation}
F = \displaystyle\frac{\left \langle {\sum_{i=1}^{N} T_i(1-T_i)}\right 
\rangle }
{\left \langle {\sum_{i=1}^{N}T_i}\right \rangle }\, \, ,
\end{equation}
where the $T_i$'s are the eigenvalues of the $t^\dagger t$ matrix, $t$
being the transmission matrix. Averaging over an interval of energy values
is performed separately for the numerator and the denominator, as 
discussed in Ref.~\cite{pmepl}. 

A conductor is said to be in the diffusive regime when its length $L$ is much
greater than the elastic mean free path $l_0$ but much smaller than the 
localization length. Since, in the presence of mode mixing, the localization 
length is approximately equal to the elastic mean free path times the number 
$N$ of propagating modes, the diffusive regime is achieved if the following 
inequality holds:
\begin{equation}
l_0 \ll L \ll N l_0\, \, . 
\label{geppo}
\end{equation}
It has been theoretically
predicted~\cite{been,nagaev} and experimentally observed in metallic
wires~\cite{henny}
that in such a condition shot noise is suppressed by a factor 1/3.
Achieving the diffusive regime in a mesoscopic conductor obtained in 
the 2DEG of a semiconductor heterostructure can be rather difficult,
because of the relatively small (up to a few hundreds) number of propagating
modes~\cite{pm-icnf11,pm-fnl}.

Chaotic cavities, characterized by constrictions that are much narrower than
the main part of the cavity, and that connect it to two reservoirs, 
are another well investigated type of mesoscopic structure and have a potential
for practical applications, such as detectors~\cite{whitney,jap}. 
In the case of a symmetric chaotic
cavity, the distribution $p(T_n)_{\rm cav}$ of the transmission eigenvalues
has a bimodal shape with peaks at $T=0$ and $T=1$. Its expression is
given by~\cite{jala}
\begin{equation}
p(T)_{\rm cav} = {\frac{1}{\pi}\frac{1}{\sqrt{T(1-T)}}}\, .
\label{distrsymm}
\end{equation}

This leads to a Fano factor equal to 1/4, a result that has also received
experimental confirmation~\cite{ober1}.

This property has been often investigated in relationship with the ratio 
of the dwell time in the cavity $\tau_D$ to the characteristic quantum 
scattering time $\tau_Q$ or Ehrenfest 
time~\cite{aleiner,suk,brouwer,whitneyshom}: 
if $\tau_D \gg \tau_Q$ quantum diffraction will have sufficient time to 
generate a quantum chaotic, noisy behavior.

In several papers the achievement of quantum chaotic dynamics has been  
attributed to the classically chaotic shape of the cavity~\cite{aleiner} 
(a stadium, for example), to the presence of impurities in the 
body of the cavity~\cite{natpap} or of
irregularities in the boundary~\cite{blanter,heusler,natpap}. 
However, it has been demonstrated by Aigner {\sl et al.}~\cite{aigner}
and by some of us~\cite{pmepl}, by means of numerical simulations, that a
Fano factor of 1/4, corresponding to the RMT result for a symmetric cavity, 
is achieved also in a cavity with a perfectly regular shape, such as a 
rectangle, 
with a completely flat potential inside, as long as the conditions on symmetry 
and on the width of the constrictions (which must be narrow enough to 
guarantee the presence of sufficiently strong diffraction and a long enough
dwell time for the electrons) are satisfied. In a rectangular cavity 
diffraction originates from the apertures and from the corners. Since 
such features have a size comparable to the Fermi wavelength $\lambda_F$, 
the uncertainty $\delta \theta$ in the diffraction angle will 
be rather large, being given by $\delta \theta=\lambda_F/a$,
where $a$ is the size of the scatterer~\cite{allark}. This implies that 
a substantial divergence of the trajectories can be achieved without the
help of classical chaos.

Some authors are of the opinion that, although not necessary for the 
achievement of chaotic dynamics, impurities may contribute to approaching the
regime with a Fano factor of 1/4 if added to a cavity which, for example, 
is characterized by a lower value of the Fano factor because of too large
openings. This has been discussed by Aigner {\sl et al.}~\cite{aigner},
Rotter {\sl et al.}~\cite{rotter}, and Jacquod and Whitney~\cite{jacquod},
who, on the basis of different approaches, argue that, as the strength of the 
disorder is increased, the Fano factor for a cavity that would otherwise 
exhibit a stronger suppression of shot noise tends to 1/4.
However, in the paper by Aigner {\sl et al.}~\cite{aigner} the
authors observe a slight increase beyond the value of 1/4 for their largest 
choice of the disorder strength, but do not explore the effects of a further 
increase of such a parameter, also because they rely on an analytical 
expression (their 
Eq.~(5)) according to which the Fano factor would be given by 1/4 times the 
integral of the dwell time distribution between a modified Ehrenfest time
$\tilde \tau_E$ resulting from the contribution of diffusive scattering and 
infinity. Based on this expression, the noise suppression
factor could at most reach 1/4, as $\tilde \tau_E\rightarrow 0$. 

Also Jacquod and Whitney do not investigate a further increase of
the amplitude of the disorder to see whether the Fano factor actually saturates
to the limit of 1/4. 

A similar conclusion is reached by Sukhorukov and Bulashenko who 
argue~\cite{suk}, with a rather complex calculation based on full counting
statistics, that homogeneous disorder in the cavity leads, as the dwell time is
increased, to the Random Matrix Theory (RMT) result and therefore, for 
a cavity with symmetric constrictions, to a suppression factor of 1/4.

Here we try to gain a better understanding of the action of disorder on 
the Fano factor for a cavity, using numerical simulations, since analytical
techniques based on the ratio of the Ehrenfest time or of a modified 
Ehrenfest time to the dwell time, although very powerful and suggestive of
intuitive explanations, are difficult to apply in a rigorous fashion to 
practical situations in 
which cavity confinement and diffusive or quasi-diffusive scattering coexist. 
For our calculations, we consider the
Gallium Arsenide/Aluminum Gallium Arsenide material system, in the effective
mass approximation (assuming an effective mass for the electrons equal to 
0.067~$m_0$, where $m_0$ is the free electron mass). For other material 
systems (such as graphene) the effective mass approximation is not 
applicable, but for structures like cavities it is still possible to use an 
envelope function approximation~\cite{kp}.
 
We start out with a simple model, in which the scatterers,
mainly corresponding to impurity charges and ionized
dopants, are represented with square obstacles with a given height. We study 
the noise properties of the cavity as a function of the scatterer strength 
(which in this model corresponds to the height of the square obstacles, 
and to their concentration).

In particular, we will show that, contrary to what can be found in the
existing literature, as disorder is increased the Fano factor does not tend
to 1/4, but, rather, crosses such value without any plateau and reaches
the diffusive limit 1/3, which is also exceeded as strong localization
develops.

Then, the general validity of the results obtained from the simple model is
verified considering realistic potential fluctuations that can be found
in a mesoscopic cavity based on a GaAs/AlGaAs heterostructure, assuming 
either a hard-wall model for the cavity or a realistic potential profile, 
resulting from the electrostatic action of depletion gates at the 
heterostructure surface.

\section{Numerical method and results}
Transport and noise properties have been
investigated using a recursive Green's function technique~\cite{sols,mgr},
with a representation in real space in the longitudinal direction and
over the transverse eigenmodes in the transverse direction. The device is
subdivided into transverse slices: within each slice the potential
is assumed to be longitudinally constant, and they are initially considered 
to be isolated and with Dirichlet boundary conditions at their
ends. The Green's function of each slice can thus be analytically evaluated, in
the form of a diagonal matrix. Then, two adjacent sections are connected
via a perturbation potential $\hat{V}$, and the overall Green's function
is obtained using Dyson's equation. The recursive application of this
procedure allows us to compute the Green's function of the whole structure,
as well as, using a modified version~\cite{sols} of the Fisher-Lee 
relation~\cite{fisher,szafer}, 
the transmission matrix. Finally, from the transmission matrix, the conductance
and shot noise power spectral density are 
obtained, within the Landauer-B\"uttiker formalism.

\begin{figure}[t!]
\begin{center}
\includegraphics[width=8cm]{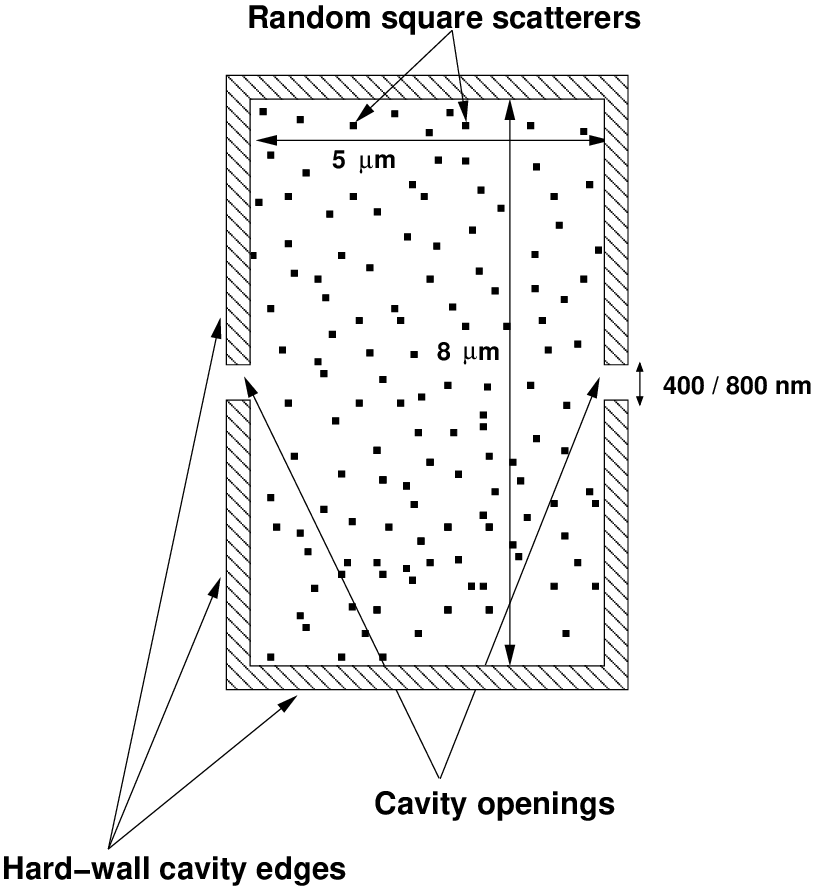}
\caption{Sketch of a 5 $\mu$m long and 8 $\mu$m
wide hard-wall mesoscopic cavity cavity with  $50\times 50$ nm$^2$ random
scatterers and 400 nm or 800 nm wide openings.}
\label{figure1}
\end{center}
\end{figure}

We start by considering a rectangular hard-wall cavity 
with a length of 5~$\mu$m and a width of 8~$\mu$m, as in the case
of the experimental device of Ref.~\cite{ober1}, with 400~nm wide openings,
containing $50 \times 50$~nm$^2$ randomly located scatterers. A sketch of the 
resulting
potential landscape is reported in Fig.~\ref{figure1}.

Results have been averaged over a set of 41 uniformly spaced energy
values, in a range of 80~$\mu$eV (corresponding to about $10 kT/e$ for a
temperature of 100~mK, with $k$ being the Boltzmann constant and $T$ the 
absolute temperature) around a Fermi energy of $E_F=9$~meV,
typical of electrons in the 2DEG of a GaAs/AlGaAs heterostructure~\cite{pmepl}.

\begin{figure}[t]
\begin{center}
\includegraphics[width=8cm]{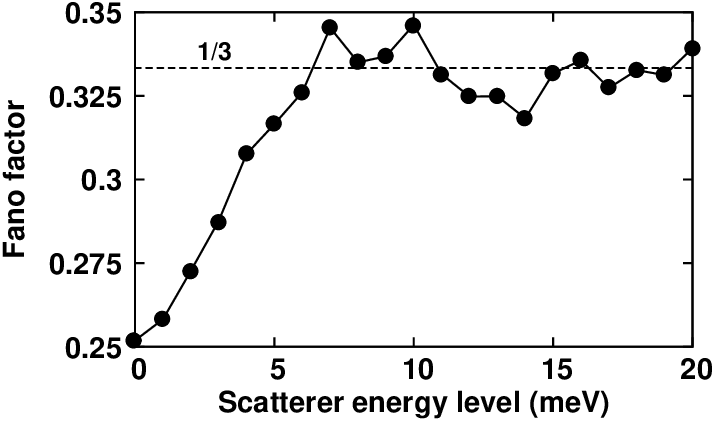}
\caption{Fano factor for a 5 $\mu$m long and 8 $\mu$m wide cavity, 
as a function of the height of 1100 square $50\times 50$~nm$^2$ square 
scatterers, for $E_f=9$~meV. 
The cavity has 400 nm wide openings, which, in the absence of randomly
located scatterers, lead to a Fano factor of $1/4$.} 
\label{figure2}
\end{center}
\end{figure}

We have initially considered
a constant number of 1100 scatterers and we have studied the behavior of the
shot noise suppression factor as a function of their height in the 0-20~meV
range. Results are reported in Fig.~\ref{figure2}: for a height of the
scatterers that is much less than the Fermi energy of the electrons, there is
substantially no significant effect and the Fano factor equals $1/4$, as in
an ideal cavity.

As the height of the scatterers is increased, the Fano factor  
approaches the value $1/3$ and seems to saturate around it as the
scatterer height grows past 10~meV. From this result one would be tempted
to conclude that, as the strength of the scatterers is increased,
the noise behavior evolves from that typical of a chaotic cavity to
that of a diffusive conductor. However, saturation to 1/3 is observed just
because, as the height of the scatterers rises above the Fermi level, their 
contribution
becomes equivalent to that of hard-wall scatterers of infinite height, and
therefore any further increase of their actual height has no effect.

\begin{figure}[t]
\begin{center}
\includegraphics[width=8cm]{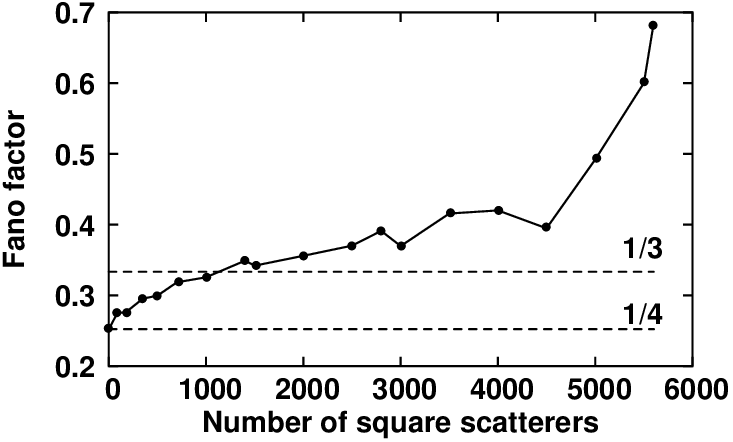}
\caption{Fano factor as a function of the
number of scatterers, for a 5 $\mu$m long and
8 $\mu$m wide mesoscopic cavity, with 400 nm wide openings and $E_F = 9 $~meV.}
\label{figure3}
\end{center}
\end{figure}

In order to vary the strength of the disorder in a way that does not lead
to saturation,
we have acted upon on the density of the scatterers. We have considered 
the same square $50\times 50$~nm$^2$ scatterers, but with a height of 15~meV 
(corresponding
to a hard-wall behavior at the considered Fermi energy) and varied their total 
number. Results are shown
in Fig.~\ref{figure3}: we notice that, as the scatterer number increases, 
the Fano factor crosses the value $1/3$ and rises well beyond it, due to 
strong localization.
In particular, the interval over which it is close to the 1/3
limit is quite narrow, and it clearly contains the previously considered value
of 1100 obstacles.
This is in agreement with the results that we have previously 
reported~\cite{pm-icnf11,pm-fnl} for a disordered conductor without any 
cavity, for
which
we have shown that it is quite unlikely that diffusive transport will be 
reached under the typical conditions that can be found for mesoscopic 
structures based on
semiconductors, mainly because of the limited number of propagating modes.

\begin{figure}[t]
\begin{center}
\includegraphics[width=8cm]{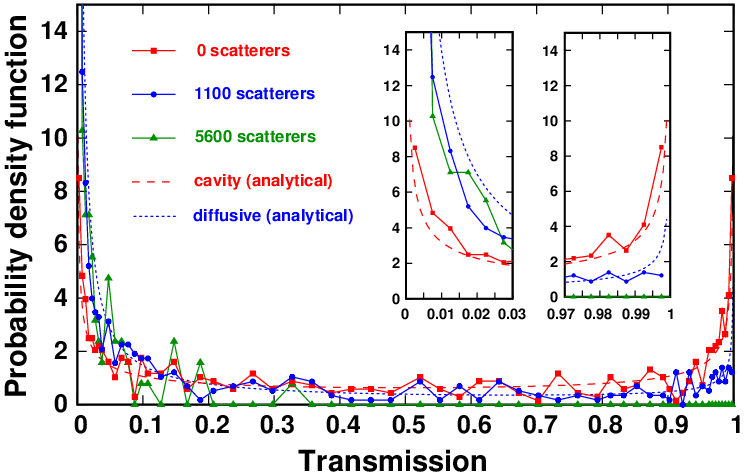}
\caption{Probability density function of the transmission eigenvalues,
for an empty cavity (squares) and for a cavity containing 1100 (circles) and 
5600 (triangles) square scatterers. For comparison, we report also the
probability density function analytically expected for an
empty cavity (dashed curve) and for a diffusive conductor (dotted curve). 
The insets contain enlargements of the regions close to zero and unitary
transmission.}
\label{figure4}
\end{center}
\end{figure}

We have also investigated the probability density function of the
transmission eigenvalues, which provides an insight into the transport
regime~\cite{blantbuett}. The interval (0,1) of possible values of the
transmission eigenvalues has been subdivided into 200 equal subintervals and  
(for 81 uniformly spaced energy levels in the considered energy range) the 
fraction of the $N$ modes
propagating through the overall structure with a transmission value
within each subinterval has been evaluated. These data, divided by
the width of the subintervals, are reported in Fig.~\ref{figure4} for an
empty cavity and for a cavity containing 1100 or 5600 square scatterers.
As can be seen, increasing the number of scatterers the fraction of
transmission eigenvalues around zero increases, while the peak around one
decreases and finally disappears.
In particular, we notice that in the absence of scatterers the probability 
density function is the one analytically expected for an empty cavity 
(i.e. that of Eq.~(\ref{distrsymm}), plotted with a dashed line), 
while when the Fano factor crosses 1/3 (for 1100 scatterers) it approaches 
the behavior 
expected~\cite{dejong3} for a diffusive 
conductor (reported with a dotted line).

For a diffusive conductor, the expected
probability density function for the transmission eigenvalues is given 
by:~\cite{dejong3}
\begin{equation}
p(T)_{\rm diff}=\frac{1}{b}\frac{1}{T \sqrt{1-T}} \Theta(T-T_0) \ ,
\end{equation}
where $\Theta(T)$ is the Heaviside step function and
\begin{equation}
T_0=\frac{4\,e^b}{(1+e^b)^2}
\end{equation}
is chosen in such a way that the integral between 0 and 1 of
$p(T)_{\rm diff}$ is equal to 1
(in the absence of the Heaviside step function it would
instead diverge). The value of the parameter $b$ has been chosen to fit
the conductance obtained with our
transport simulation for 1100 square scatterers.
In particular, the conductance in a diffusive structure is given by
\begin{equation}
G_{\rm diff}=G_0N\int_0^1 \! T p(T)_{\rm diff} \,dT=
G_0 N \frac{2}{b}\tanh\left(\frac{b}{2}\right)
\end{equation}
(where $G_0=2e^2/h$ is the conductance quantum) and, since from our
simulations with 1100 scatterers we have obtained $G\approx 3.89\,G_0$, we have
selected the value $b=7.2$ for the representation of the analytical curve.

\begin{figure}[t!]
\begin{center}
\includegraphics[width=8cm]{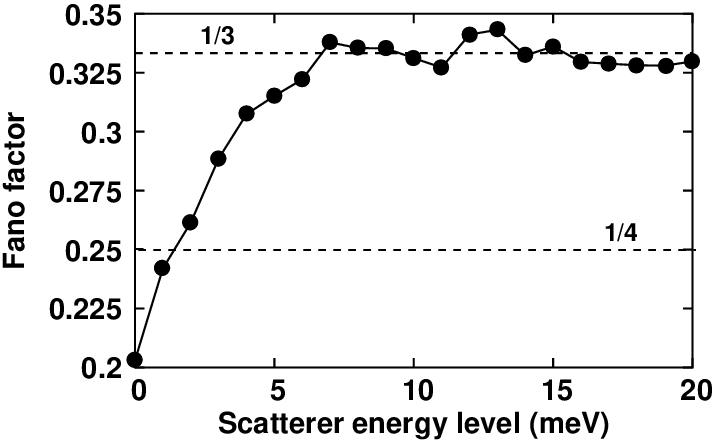}
\caption{Fano factor as a function of
energy for a 5 $\mu$m
long and 8 $\mu$m
wide mesoscopic cavity, with 800 nm wide openings and $E_F = 9$~meV,
in the presence of 1200 scatterers.}
\label{figure5}
\end{center}
\end{figure}

So far we have studied the effect of disorder on a cavity that, when clean,
already exhibits a Fano factor of $1/4$, due to symmetry and to narrow enough
constrictions. Let us now
move to a situation analogous to the one investigated in 
Refs.~\cite{rotter,jacquod}, in which we start from a clean cavity with 
wider constrictions (800~nm) and, therefore, with a Fano factor below $1/4$. 

In Fig.~\ref{figure5} the resulting Fano factor is reported as a function 
of the scatterer height for 1200 scatterers. For
very low values of the obstacle height, the Fano factor is below $1/4$,
while, when the scatterer height is increased, it crosses the value $1/4$ and
approaches the value $1/3$, as in the previously discussed case of 
Fig.~\ref{figure2}. 

\begin{figure}[t!]
\begin{center}
\includegraphics[width=8cm]{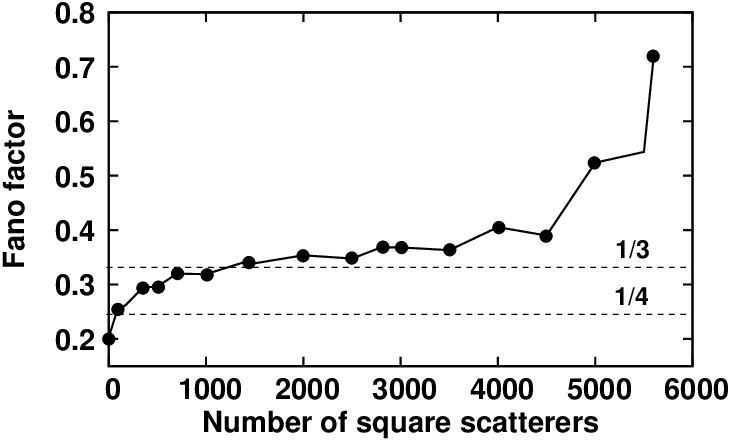}
\caption{Fano factor as a function of the
number of scatterers, for a 5 $\mu$m long and
8 $\mu$m
wide mesoscopic cavity, with 800 nm wide openings and $E_F = 9 $~meV.}
\label{figure6}
\end{center}
\end{figure}

If the number of scatterers is increased beyond $1200$, the Fano factor 
grows above $1/3$, as in the case of the cavity with 400~nm wide 
constrictions. In Fig.~\ref{figure6} the shot noise suppression factor 
is plotted as a function of the number of 
scatterers. 

These results confirm a few main points: a) the Fano factor of $1/4$ is 
achieved independent of the chaotic or regular shape of the cavity,
as long as the constrictions are symmetric and narrow enough; b) no disorder
is needed to achieve the Fano factor of $1/4$; c) in the presence
of disorder the Fano factor rises: if originally below $1/4$ crosses the 
value $1/4$ without any appreciable plateau, and reaches, as the disorder
strength is increased, the diffusive limit of $1/3$, which is crossed, too,
for choices of the parameters typical of mesoscopic semiconductor structures,
with an increase towards 1, as the strong localization regime is approached.

\begin{figure}[t!]
\begin{center}
\includegraphics[width=8cm]{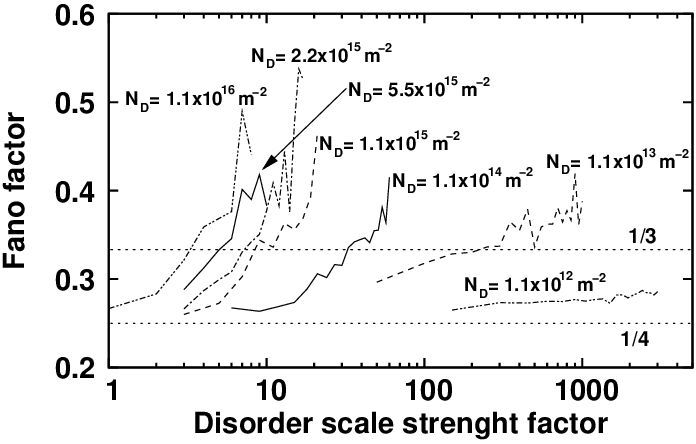}
\caption{Fano factor for a 5 $\mu$m long and 8 $\mu$m
wide hard-wall mesoscopic cavity, with 400 nm wide openings, as a
function of the scale factor M multiplying the realistic potential
fluctuation,
for 7 different impurity concentrations $N_D$ and for $E_F = 9 $~meV.}
\label{figure7}
\end{center}
\end{figure}

In order to understand whether the results obtained so far are of general 
validity, we have studied the effect of a more realistic potential profile
on shot noise suppression. We have started by considering the same 5 $\mu$m 
long and 8 $\mu$m wide hard-wall mesoscopic cavity with 400 nm wide openings, 
with the addition of potential fluctuations at the level of the 2DEG resulting
from randomly located ionized dopants and charged impurities, which, at low
temperatures, represent the dominant source of scattering. 

In order to evaluate their contribution to the potential landscape at the
2DEG level, we have resorted to the semi-analytical
expression given by Stern and Howard~\cite{stern} for the screened
potential due to a point charge located at a distance $D$ from the 2DEG.

We have initially computed a potential profile resulting
from ionized impurities located $D=40$~nm away from the 2DEG,
which is a realistic distance for actual heterostructures, where a spacer
layer separates the dopants from the 2DEG, in order to reduce their effect on
transport.

Even without a fully self-consistent approach, the generation of such a 
potential is a computationally intensive task, due to the relatively large 
area of the cavity. Therefore, in order to keep the computational complexity 
under control, the full computation has been performed only for $D=40$~nm, 
and the disorder strength has then been varied by simply scaling the result 
by a factor $M$, instead of repeating the full calculation for different
values of $D$.

Since we are considering dopants distributed over a finite region, this leads
to a fictitious minimum of the potential in the middle of the structure,
due to the reduced number of dopants acting upon points in the region
near the boundaries~\cite{jap}. In order to avoid this effect and to keep
a flat average potential, we have resorted to the approximate approach
mentioned in Ref.~\cite{pm-icnf11}, including also positive impurities, in a
number equal to the negative ones.

With this approach, the behavior of
the Fano factor has been computed for 7 different impurity densities $N_D$, 
as a function of the disorder strength scale factor $M$. 

Results are shown in Fig.~\ref{figure7}: as we can see, for impurity 
concentrations that are typical of practical heterostructures 
($N_D \approx 10^{15}$~m$^{-2}$) the Fano factor,
which is close to $1/4$ for very low values of the parameter $M$
as expected, rapidly increases to values well above 1/3. 

As the disorder strength is raised, the curves in Fig.~\ref{figure7} are
characterized by larger fluctuations, because the number of modes with 
significant transmission decreases, and the energy averaging that we 
perform is not sufficient to achieve a smooth curve any longer. To make 
it smoother we should 
perform an ensemble average over different realizations of the random
potential landscape, which however would not be very meaningful, 
since it would not correspond to what can be measured in an actual experiment
(except for the unlikely case in which many cavities were connected together
in parallel and measured at the same time). 

If we compare
these results with those obtained in the absence of the mesoscopic cavity, for
a purely disordered conductor~\cite{pm-icnf11},
we see that, although in the present case the behavior is less smooth and 
the region of diffusive transport is narrower, the Fano factor crosses the 
value $1/3$
for about the same impurity strength.
This indicates that, also in the presence of more realistic disorder
inside the cavity, the noise properties of the whole system are dominated
by the diffusive (or better, for most values of the disorder strength, 
quasi-diffusive) regions. 
Moreover, since the fully diffusive regime is achieved within
a very narrow interval of the disorder strength scale factor $M$, we
observe only a rapid transition of the Fano factor through the value $1/3$.
This behavior is very similar to what we have previously observed
in the presence of hard-wall scatterers.

Something closer to a plateau is observed only for  very low impurity 
concentrations ($N_D=1.1 \times 10^{13}$~m$^{-2}$) and very large values
of the parameter $M$, as shown in Fig.~\ref{figure7}. However, as
previously demonstrated~\cite{pm-icnf11}, with this choice of parameters
the resulting potential fluctuations appear definitely unrealistic for a
semiconductor-based device. Indeed, there would be few very large, in terms
of their extension on the 2DEG plane, scatterers,
while in GaAs/AlGaAs heterostructures typical
impurities cause very sharp and localized potential fluctuations.

Moreover, in Fig.~\ref{figure7} we can observe that for the lowest considered
impurity density ($N_D=1.1\times 10^{12}$~m$^{-2}$), the $1/3$ value
is not reached even for very large values of the scale factor $M$, 
which indicates that transport becomes almost ballistic.

\begin{figure}[t!]
\begin{center}
\includegraphics[width=8cm]{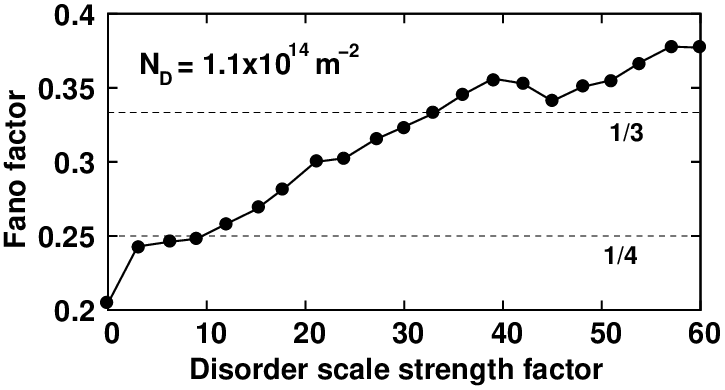}
\caption{Fano factor as a function of the
disorder strength scale factor $M$, for a 5 $\mu$m long and 8 $\mu$m
wide mesoscopic cavity, with 800~nm wide openings. The Fermi energy is $9$~meV
and the donor density is $1.1 \times 10^{14}$~m$^{-2}$.}
\label{figure8}
\end{center}
\end{figure}

Then, also for this type of disorder we have considered a cavity with wider
constrictions (800~nm wide, as in the case 
of Fig.~\ref{figure6}), assuming a donor density 
$N_D=1.1 \times 10^{14}$~m$^{-2}$. 
The results for the Fano factor as a function of the disorder strength
are reported in Fig.~\ref{figure8}: we observe a behavior
analogous to that for the case of hard-wall scatterers, with the 
shot noise suppression factor increasing from a value below 1/4 (since for
a null disorder scale strength factor, i.e. for a clean cavity, the 
width of the constriction is not small enough to achieve
the regime characterized by a Fano factor of $1/4$), crossing $1/4$, reaching 
$1/3$ and increasing beyond $1/3$. 

If we again compare the results reported here with those obtained in the 
absence of the constrictions defining the cavity (see Ref.~\cite{pm-icnf11}), 
we notice that
the value of $1/3$ is reached for approximately the same parameter 
values ($N_D$ and disorder scale strength factor),
i.e. in the narrow interval within which the conductance satisfies the 
condition of Eq.~(\ref{geppo}) for diffusive transport. 
This confirms that in conductors with strong disorder the noise properties
become almost completely independent of the presence of the constrictions that 
define the cavity. 

\begin{figure}[t!]
\begin{center}
\includegraphics[width=7cm]{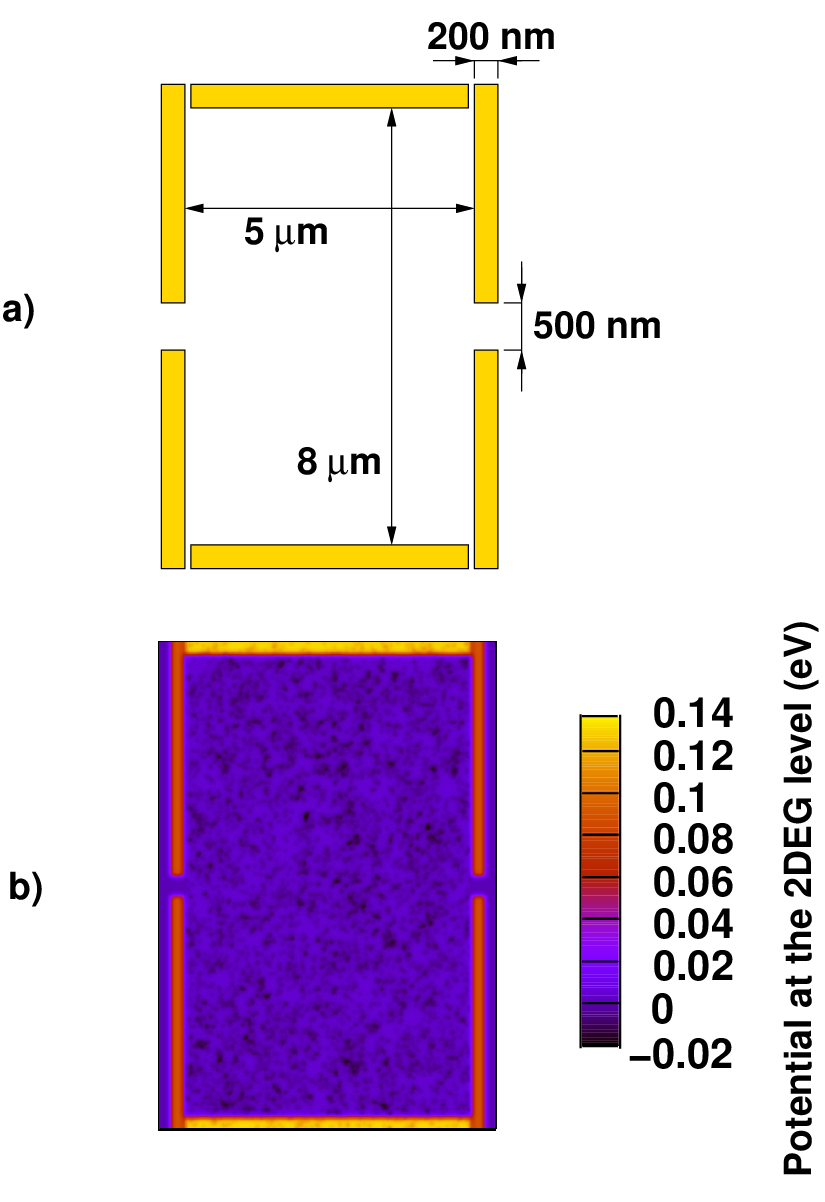}
\caption{Upper panel: layout of the depletion gates defining the device;
lower panel: electrostatic potential energy at the 2DEG level
for a bias
of $-3.5$~V applied to the horizontal gates and $-2.5$~V to the split
gates, in the presence of fluctuations due to randomly located dopants.}
\label{figure9}
\end{center}
\end{figure}

Finally, we have studied a device that is realistic also from the point of 
view of the confining potential, obtained by depletion gates deposited on
the heterostructure surface. Such a potential is included 
considering the screening effect of the 2DEG, which provides an essential 
contribution.

The solution of the complete self-consistent problem for
a device with a size of several micrometers would be very demanding from the
computational point of view. Therefore, for the evaluation of the potential
profile at the 2DEG level, we have used a semi-analytical approach proposed
by Davies {\em et al.}~\cite{davies}. We can consider
the two-dimensional electron gas as an equipotential surface when
its kinetic energy is much smaller than the electrostatic potential
between the surface and the 2DEG. In this case, we can solve the Poisson
equation with Dirichlet boundary conditions: at the surface we consider
the potential equal to the gate bias under each electrode and zero
elsewhere, and at the 2DEG level the potential is assumed to be constant and
equal to zero. Then, combining Coulomb's theorem and the expression for the
two-dimensional density of states, we obtain that the screened potential,
to first order, is given by
\begin{equation}
\phi_{scr}=-\frac{\pi \hbar^2}{m^*}\frac{\varepsilon_0 \varepsilon_r}{e^2}
\frac{\partial \phi}{\partial z}\bigg|_{z=d}\,
\end{equation}
with $\varepsilon_0$ being the vacuum permittivity, $\varepsilon_r$ the
relative permittivity of the semiconductor, $\phi$ the electrostatic
potential, $d$ the depth of the 2DEG with respect to the surface,
$m^*$ the electron effective mass, and
$\hbar$ the reduced Planck constant.

For the potential fluctuations inside the cavity we have used the same 
approach that has been previously described.

In the upper panel of Fig.~\ref{figure9} the layout of the gates that create
the cavity is shown. The split gates defining the constrictions have been
designed in such a way as to have the same number of propagating modes as
in the hard-wall cavity with 400~nm wide constrictions; this requires 
a gap of 500~nm, with a bias of $-2.5$~V. The bias applied to the horizontal
gates is instead of $-3.5$~V. 
In the lower panel of Fig.~\ref{figure9} we show the resulting potential at the 
2DEG level, with the inclusion of disorder.

\begin{figure}[t!]
\begin{center}
\includegraphics[width=8cm]{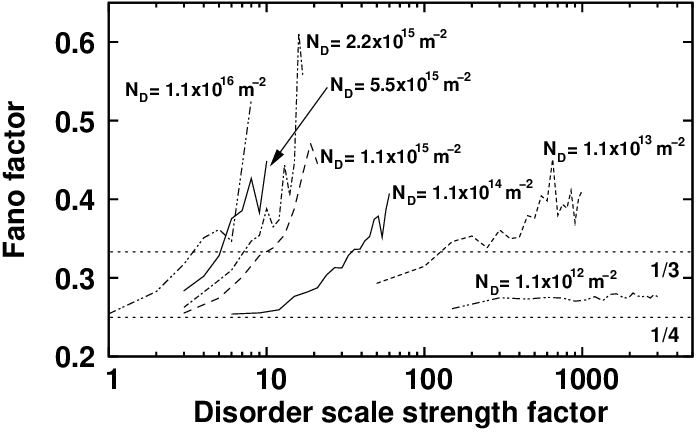}
\caption{Fano factor for a 5 $\mu$m long and 8 $\mu$m
wide mesoscopic cavity with realistic confinement potential, defined by 
split gates with a gap of 500~nm, 
as a function of the scale factor $M$ multiplying the amplitude of the 
disorder,
for 7 different impurity concentrations $N_D$ and for $E_F = 9 $~meV.}
\label{figure10}
\end{center}
\end{figure}

In Fig.~\ref{figure10} the Fano factor as a function of the disorder scale
factor $M$ for 7 different impurity densities $N_D$ is shown. The
results are comparable to what we have observed in Fig.~\ref{figure7}. This
indicates that the specific details of the potential profile defining the 
cavity do not play an essential role in determining the noise properties 
of the whole system. Also in this case, for typical $N_D$ values of 
realistic heterostructures, the Fano factor rises rapidly from $1/4$ to 
values well above $1/3$, and a behavior close to the diffusive one is observed 
only in the case of low impurity concentrations
and unrealistically large values of $M$. Once again, for $N_D=1.1\times
10^{12}$~m$^{-2}$ we can argue that transport is almost ballistic.

\bigskip

\section{Conclusion}
We have performed a detailed numerical analysis of the effect of disorder on 
shot noise suppression in mesoscopic cavities, with the aim of reaching
general conclusions and clarifying issues that had been left open in the 
existing literature.

Our results, while further confirming that the characteristic suppression
down to 1/4 of full shot noise in symmetric mesoscopic
cavities can be achieved in cavities with a regular shape and as a result
of diffraction due just to the constrictions and to the corners,
makes clear that the inclusion of disorder leads to
an increase of the Fano factor, and, as its strength is raised, 
a shot noise behavior typical of disordered conductors is reached, 
independent of the presence of the constrictions. 

In particular, for cavities with 
constrictions that are narrow enough to yield, without disorder, a Fano
factor of 1/4, the addition of disorder leads to an increase of shot noise
up to 1/3 and beyond, with a plateau around 1/3 only for 
choices of the parameters that warrant the achievement of the diffusive
regime in a range of values of the disorder strength.
In the case of cavities with wider constrictions, the Fano factor in the 
absence of disorder is below 1/4 and, as the disorder strength is increased,
raises, crossing 1/4 without any visible plateau and then evolving in a way 
analogous to that of the cavities with narrower constrictions. This confirms
that disorder does not contribute towards the achievement of the chaotic
regime associated with the Fano factor of 1/4 in the way predicted by the 
existing literature.
 
In the case of strong disorder, the body of the cavity acts as
a diffusive or quasi-diffusive conductor that 
effectively separates the entrance and exit constrictions and dominates
the noise behavior.

Furthermore, from our results it is apparent that the overall noise 
characteristics are not strongly dependent on the details of the confinement 
potential and on the nature of the specific disorder being considered.

\bibliographystyle{apsrev}

\end{document}